\documentclass[twocolumn,secnumarabic,amssymb, nobibnotes, aps, prl, superscriptaddress]{revtex4-1}

\usepackage{float}
\usepackage{hyperref, url}
\usepackage{graphicx}
\usepackage{amsmath} 
\usepackage{epstopdf}
\usepackage{setspace}
\usepackage[shortlabels]{enumitem}
\usepackage{mathrsfs}
\usepackage{color,soul}
\usepackage[dvipsnames]{xcolor}
\usepackage{dcolumn,amssymb,subfig}
\usepackage{comment}
\usepackage{caption}

\usepackage{dcolumn} 
\usepackage{natbib}
\usepackage{bm} 
\usepackage{float} 
\usepackage{bbm} 
\usepackage{color}
\usepackage{amsfonts}
\usepackage{lmodern} 
\usepackage{amsmath,amssymb}
\usepackage[normalem]{ulem}
\renewcommand{\vec}[1]{\boldsymbol{#1}}

\begin{document}

\title{Quadrupolar active stress induces exotic phases of defect motion in active nematics}

\author{Salik A. Sultan}
\thanks{These authors contributed equally to this work.}
\affiliation{The Niels Bohr Institute, University of Copenhagen, Copenhagen, Denmark}
\author{Mehrana R. Nejad}
\thanks{These authors contributed equally to this work.}
\affiliation{Rudolf Peierls Centre for Theoretical Physics, University of Oxford, UK}
\author{Amin Doostmohammadi}
\thanks{corresponding author: doostmohammadi@nbi.ku.dk}
\affiliation{The Niels Bohr Institute, University of Copenhagen, Copenhagen, Denmark}

\begin{abstract}
A wide range of living and artificial active matter exists in close contact with substrates and under strong confinement, where in addition to dipolar active stresses, quadrupolar active stresses can become important. Here, we numerically investigate the impact of quadrupolar non-equilibrium forces on the emergent patterns of self-organisation in non-momentum conserving active nematics. Our results reveal that beyond having stabilising effects, the quadrupolar active forces can induce various modes of topological defect motion in active nematics. In particular, we find the emergence of both polar and nematic ordering of the defects, as well as new phases of self-organisation that comprise topological defect chains and topological defect asters. The results contribute to further understanding of emergent patterns of collective motion and non-equilibrium self-organisation in active matter.
\end{abstract}
\maketitle

\section{Introduction}
Active materials are out of equilibrium systems in which energy is continuously consumed and transferred to mechanical work by constituent elements of the matter \cite{sanchez2012spontaneous,weirich2019self,Marchetti13,needleman2017active,julicher2018hydrodynamic}.
Examples, at various length scales, include
flocks of birds \cite{tonertu1998flock}, schools of fish, networks of actin and microtubules \cite{bate2019microtubule} and cellular tissues \cite{duclos2018spontaneous, kawaguchi2017topological}.
In contrast to most non-equilibrium systems, where an external driving force is applied to the system, the defining attribute of active materials is that the energy is input locally at the level of individual components of a structure, leading the system into a dynamic state. The local energy injection results in the generation of active stresses by the particles that in turn drive distinct patterns of collective motion and self-organisation of active materials. Examples range from large-scale coordinate motion of cells~\cite{poujade2007} and bacteria~\cite{meacock2020bacteria}, to coherent flows of subcellular filaments under geometric confinement~\cite{wu2017}. 

The active stresses, therefore, are often considered to induce destabilising effects, putting an otherwise quiescent matter into motion~\cite{ramaswamy2002hydrodynamics,voituriez2005spontaneous,thampi2015intrinsic,giordano2021activity}. The destabilising effects of active stresses can even lead to chaotic flows of active particles in a state termed active turbulence~\cite{giomi2011patterns, thampi2016activeturbulence}. However, because of their widespread implications in biological processes and in the design of non-equilibrium materials, there is a growing interest in control and stabilisation of active materials~\cite{needleman2017active,doostmohammadi2018active,brusatin2018biomaterials}. In this venue, several theoretical and experimental mechanisms have been proposed to tame, the otherwise chaotic motion of active particles~\cite{doostmohammadi2019ceilidh,shankar2021topological}. Particularly successful approaches use topological~\cite{keber2014topology,metselaar2019topology,carenza2020chaotic} and geometric constraints~\cite{hardouin2019reconfigurable} to streamline flows of active particles. Mechanistically, such constraints induce hydrodynamic screening effects that allows for stabilisation of active flows. For example, placing bacterial suspensions or microtubule-motor protein mixtures under confinement is shown to result in a crossover from chaotic flows to vortex-lattices and coherent streams as the confinement size decreases~\cite{wioland2016directed,hardouin2019reconfigurable}. Similarly, placing active particles in contact with substrates results in friction-induced screening length that could stabilise chaotic flows into vortex-lattices~\cite{doostmohammadi2016stabilization,thijssen2020friction}. 

Not only do these screening effects extract momentum from the system, they also affect the dominant forms of active stresses that are exerted by particles on their surroundings. In the absence of screening effects, it is well-established that the dominant stress contribution from a mirco-scale active particle is a stresslet ~\cite{mak2016tension,ramaswamy2010mechanics,prost2015gel}. However, when an active particle is in contact with a substrate or under extreme confinement, the dipolar stress is screened and a higher order quadrupolar contribution also becomes important~\cite{arnoldmathijssen2016}. In this vein, a recent study has introduced the coarse-grained continuum representation of quadrupolar forces as an additional non-equilibrium active force that is generated by dense suspension of active particles~\cite{maitra2018nonequilibrium}. Interestingly, using linear stability analyses, it is shown that such quadrupolar force can induce a stabilising effect on the active system, counteracting the hydrodynamic instability that is induced by the dipolar forces. Notwithstanding this important contribution, the non-linear impacts of such additional non-equilibrium force on the pattern formation and dynamic organisation of active systems remain unexplored. Going beyond the linear instability, here we numerically examine the full non-linear effects of the additional quadrupolar force on the dynamics of collective pattern formation in a model of active nematics. We focus particularly on the dynamics of the topological defects, singularities in the orientation field, that are increasingly identified in various realisations of active matter and are shown to have biological functions~\cite{Saw2017,kawaguchi2017topological,copenhagen2020bacteria,meacock2020bacteria,maroudas2020topological,zhang2021defects,blanchmercader2021topological,doostmohammadi2021physics}. The results uncover exotic phases of defect organisation that comprise topological defect chains, asters of topological defects, as well as polar and nematic ordering of the motile topological defects.
\\
 \begin{figure*}[t] 
    \centering
    \includegraphics[width=0.99\textwidth]{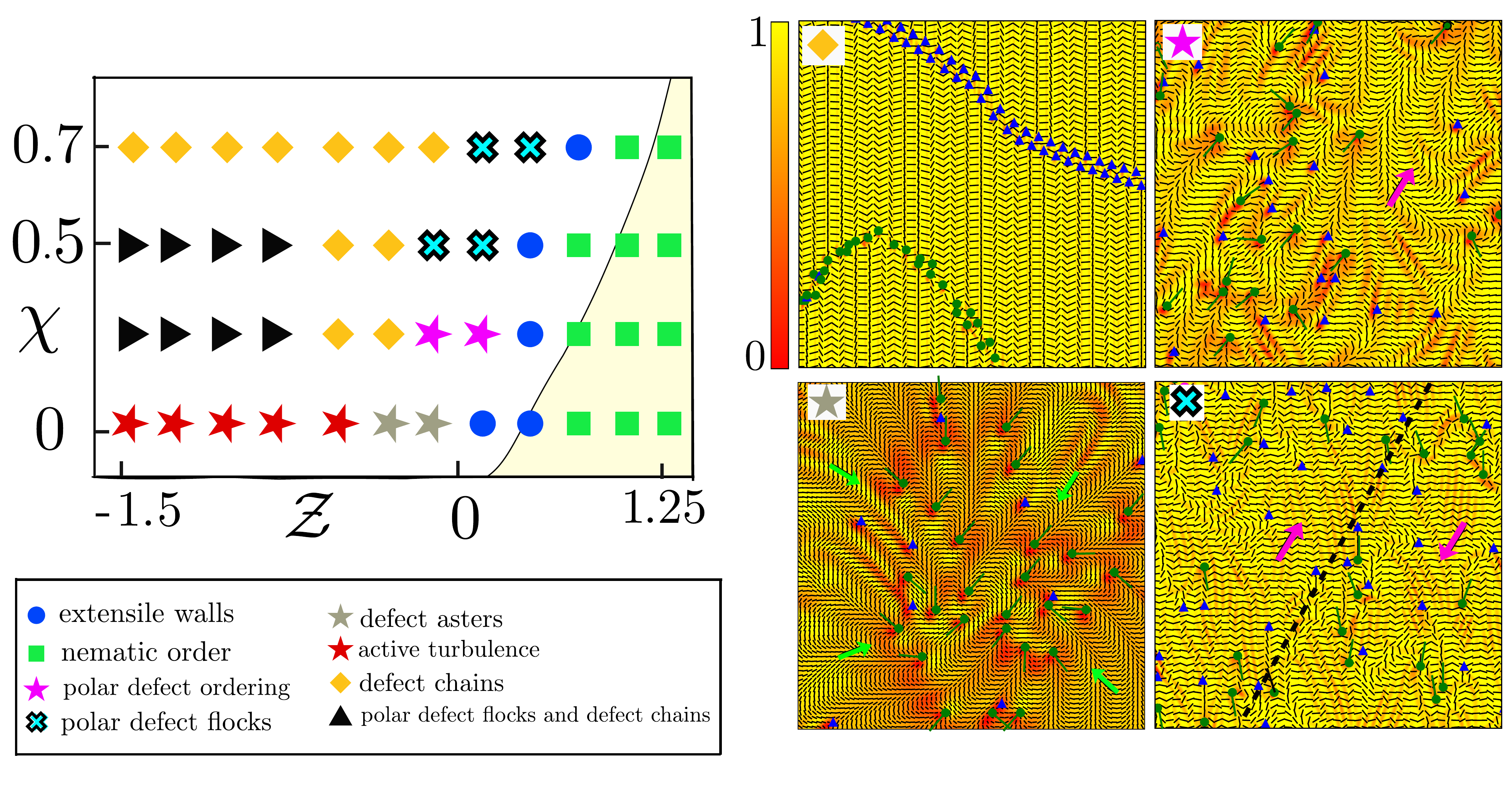}
    \caption{Phase space of the system for different values of the tumbling parameter $\chi$ and the activity ratio $\mathcal{Z}=\zeta_2/\zeta_1$. The nematic order can become stable for positive values of the force quadrupole term. The yellow region in the phase diagram shows the stable region found analytically in Ref. \cite{maitra2018nonequilibrium}. Snapshots showing the distinguishable characterization of defect chains, polar defect ordering, defect asters, polar defect flocks. The underlying color map shows the magnitude of the nematic order, and $+1/2$ ($-1/2$) defects are represented in green (blue).}
\label{fig:lsa}
\end{figure*}
The paper is organised as follows: in section~\ref{sec:sec1} we introduce our model and discuss the details of the governing equations, active forces, and provide the list of simulation parameters. In section~\ref{sec2}, we introduce different phases that are revealed through numerical simulations of the system and provide a phase diagram that represents the emergent phases as a function of dipolar and quadrupolar force strengths and the flow-aligning parameter. In section~\ref{sec3} we represent a quantitative characterisation of observed phases and analyse emergent topological defect organisations. Finally, concluding remarks and future directions are discussed.

\section{Methods\label{sec:sec1}}

In order to understand mechanisms behind diverse behaviours of active materials, it is often useful to divide active systems into two general categories: wet and dry active matter. In wet active matter, flows created by the particles mediate long-range hydrodynamic interactions in the system \cite{koch2011collective} such as in unconfined suspensions of filaments and motor proteins~\cite{needleman2017active, kumar2018tunable}, bacterial suspensions in low concentrations~\cite{turiv2020jets}, and collection of artificial active colloids suspended within a fluid medium~\cite{buerle2020formation}.
On the other hand, the collective behaviour of dry active materials is governed by direct interactions, such as collision between active particles \cite{scholz2018rotating,chate2019dry}. Shaken granular matter \cite{junot2017active,kumar2014flocking} and bacteria moving on a substrate~\cite{meacock2020bacteria} are examples of dry active matter. Descriptions based on dry active matter theories have been helpful in investigating the different patterns in clusters of bacteria, which live on dry surfaces in tight spaces \cite{puranai2013collective,grossmann2016mesoscale}. Since in this work our focus is on the behavior of active model system in the presence of strong hydrodynamic screening, we employ equations of dry active nematics~\cite{doostmohammadi2016stabilization} with the additional quadrupolar active forces. The active nematic model is chosen since it has proven successful in describing a wide variety of active materials that constitute elongated building blocks such as rod-shaped bacteria~\cite{Nagai2018selfpropelled}, subcellular filaments~\cite{alvarado2014polymers}, and spindle-shaped cells~\cite{duclos2017topological}, as well as deformable cells in which orientational ordering is an emergent feature~\cite{mueller2019monolayer, Saw2017,blanch2018turbulent}.

To describe the orientational order, we study the evolution of the nematic order parameter $\textbf{Q}=2q (\hat{n} \hat{n}-\frac{\textbf{I}}{2})$, where $q$ and $\hat{n}$ show the magnitude and orientation of the order, respectively, and $\textbf{I}$ represents the identity tensor.
\\
The dynamics of the $\textbf{Q}$-tensor is coupled to the velocity of the surrounding fluid $\vec{u}$, and reads:
\begin{align}
\frac{D Q_{ij}}{Dt} &= \chi E_{ij} + \Gamma H_{ij},\label{eq1}\\
f u_i &= -\zeta_1 \ \partial_{j} Q_{ij} - \zeta_2 Q_{ij} \partial_{k}Q_{kj}, \label{eq2}
\end{align}
 where we have introduced the convected co-rotational time derivative $ D Q_{ij}/Dt=(\partial_t + u_k \partial_k) \ Q_{ij} - \omega_{i k} Q_{k j} + Q_{i k} \omega_{k j}$, in which $\omega_{ij} = (\partial_j u_i - \partial_i u_j)/2$ is the vorticity tensor. In Eq.~\ref{eq1}, $\chi$ is the tumbling parameter and $E_{ij}= (\partial_i u_j+\partial_j u_i)/2$ is the strain rate tensor. $\Gamma$ shows the rotational diffusivity which along with $H_{ij}$, the molecular field,
describes the relaxation of the \textbf{Q}-tensor towards the minimum of a free energy described by: 
\begin{equation} \label{eqn:freef}
    \mathcal{F} = \frac{A}{2} (q^2 - \frac{1}{2}\textbf{Q}:\textbf{Q})^2+\frac{K}{2}|\nabla \textbf{Q}|^2+g (\nabla \cdot[\nabla \cdot \textbf{Q}])^2.
\end{equation}
The free energy includes a nematic alignment term (with coefficient $A$) and an elastic term which penalises gradients in the \textbf{Q} tensor. Here, we use a single elastic constant approximation and show the elastic constant by $K$. In line with previous studies of dry active nematic~\cite{oza2016antipolar,putzig2016instabilities}, we also include a regularization term with coefficient $g$ which provides
stability to our numerical solutions at small length scales. 

In Eq.~\ref{eq2}, $f$ is friction coefficient, and $\zeta_1$ and $\zeta_2$ show activity coefficient related to a force dipole and a force quadrupole, respectively. Two well-known classes of active suspensions, namely extensile and contractile systems, are described by $\zeta_1>0$ and $\zeta_1<0$, respectively. Extensile systems comprise elements that pump the surrounding fluid in an outwards direction along their symmetry axis $\hat{n}$ and inwards in the direction perpendicular to $\hat{n}$. Contrarily, in a contractile system, elements pump fluid inwards along $\hat{n}$ and outwards in a direction perpendicular to that. As a result of the flows created in active suspension in the bulk, an extensile system is stable under splay deformation and unstable under bend deformation~\cite{ramaswamy2002hydrodynamics}. Contrarily, a contractile system is stable under bend deformations, but unstable as a result of splay deformations \cite{thampi2016activeturbulence}. In addition to this well-established dipolar activity, the force distribution of the higher angular symmetry, the force quadrupole term with coefficient $\zeta_2$ in Eq.~\ref{eq2}, becomes important in systems in contact with a substrate, where the momentum is not conserved~\cite{maitra2018nonequilibrium}.

We study the role of the quadrupole active force in the dynamics of the system, by numerically solving the coupled equations of the nematic tensor $\textbf{Q}$ and the velocity field $\vec{u}$. In this regard, we fix the value of the dipolar active force to $\zeta_1=0.2$, and study the role of the tumbling parameter $\chi$ as well as the strength of the active force quadrupole $\zeta_2$ in the dynamics. As such, we vary the strength of the active quadrupole term in the range $ -0.3 \leq \zeta_2 \leq 0.25$, effectively exploring the interplay between the dimensionless ratio $\mathcal{Z}=\zeta_2/\zeta_1$ and the tumbling parameter $\chi$, which determines whether the system is in a flow tumbling or flow-aligned regime \cite{ternet1999flowaligning}. In a flow tumbling regime, the director tumbles under shear flow while in a flow aligning regime it aligns with an angle (Leslie angle) with the direction of the shear~\cite{BerisBook}. The system size of $512\times512$ grid points is simulated with periodic boundary conditions on all sides.
Parameters for the rotational diffusivity, regularisation, and friction are fixed at $\Gamma = 0.05$, $g = 0.1$, and $ \textit{f} = 10$. In addition, the elasticity, and the bulk nematic alignment coefficient are set to $\text{K} = 0.15$ and $\text{A} = 0.5$, respectively.
\section{Results}\label{sec2}
\begin{figure}[t]
      \centering
      \includegraphics[width=\linewidth]{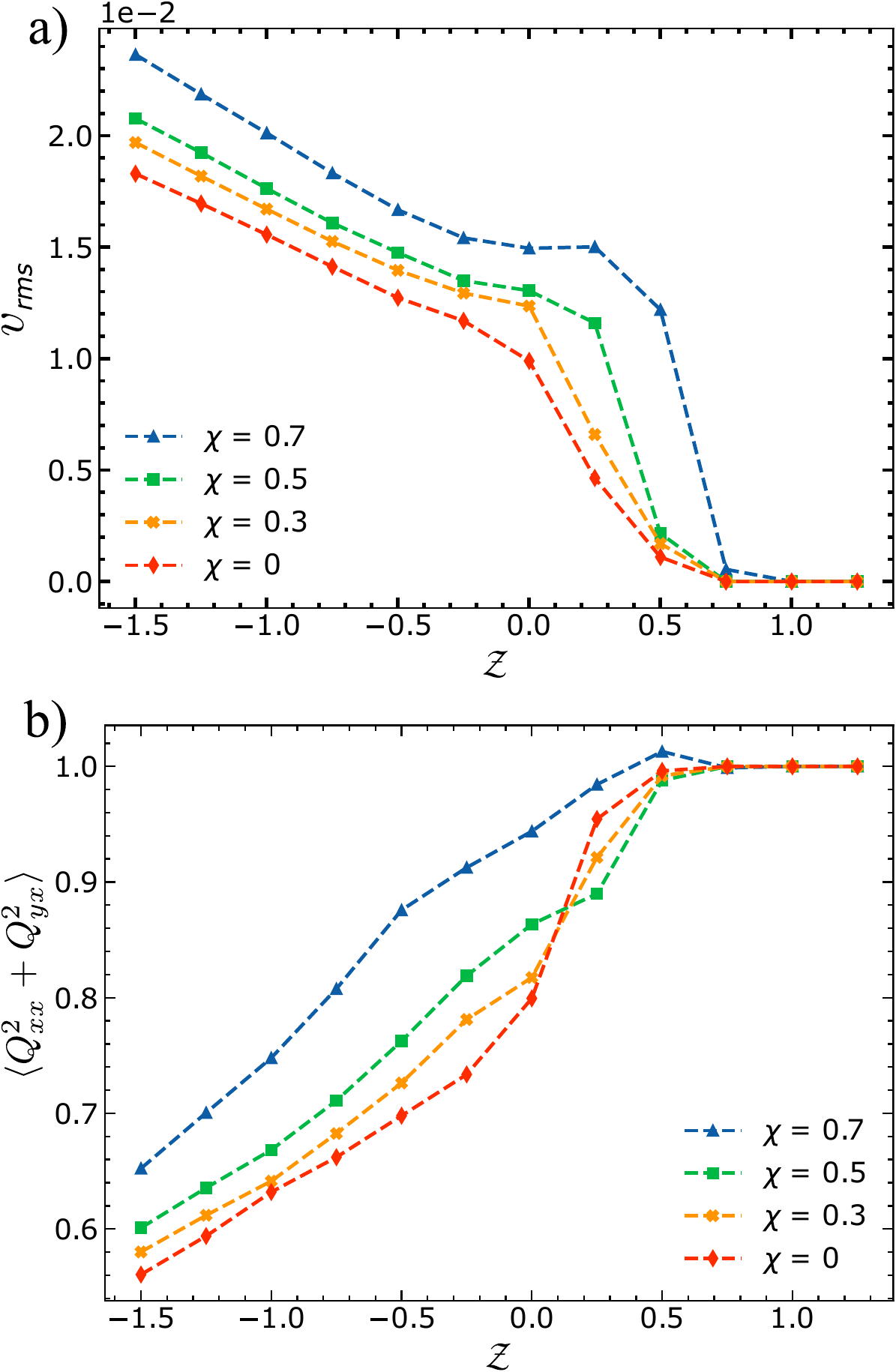}
\caption{a) Root mean square velocity $v_\text{rms}$ averaged over space and time in steady state. Regardless of flow-aligning or tumbling state $v_\text{rms}$ decreases as $\mathcal{Z}$ increases. By proxy, it is also apparent that increasing the quadrupolar active term, lowers the velocity of the system. b) Magnitude of the order defined as $\langle Q_{xx}^2+Q_{yx}^2 \rangle$ averaged over space and time in steady state. The average order follows the inverse trend compared to the rms-velocity, and increases by increasing $\mathcal{Z}$.} 
\label{rmsvel}
\end{figure}
\begin{figure}
     \centering
         \includegraphics[width=\linewidth]{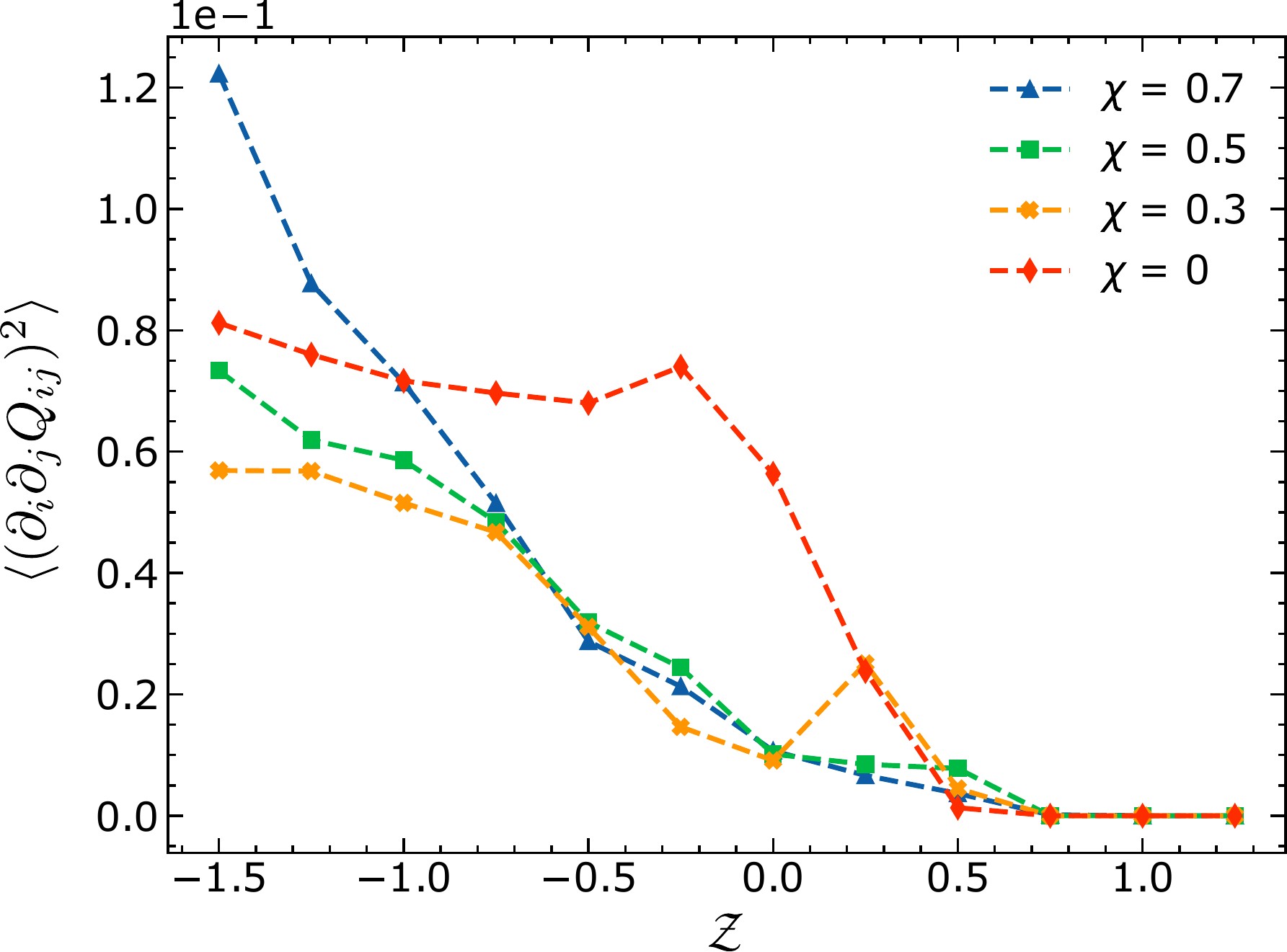}
     \caption{Average deformation defined as $\langle\Delta n \rangle=\langle(\partial_i\partial_jQ_{ij})^2\rangle$. The average is taken over time and space in steady state  The fall in deformation correlates with observed stabilisation as a consequence of increasing $\mathcal{Z}$.}
\label{deform}
\end{figure}

\subsection{Stability diagram}
We begin by investigating the emergent patterns for varying values of the quadrupolar to dipolar activity ratio $\mathcal{Z}=\zeta_2/\zeta_1$ and the flow-aligning parameter $\chi$.
The results are summarised in a stability diagram in Fig.~\ref{fig:lsa}, where distinct patterns of motion are represented.

Fig.~\ref{fig:lsa} shows that a positive value of the quadrupolar activity can stabilise the nematic phase and the required quadrupolar force for stability increases with increasing the value of the flow-aligning parameter. This is in agreement with the analytical result of Ref.~\cite{maitra2018nonequilibrium}: in Fig.~\ref{fig:lsa} we show the stable nematic phase predicted analytically in yellow background color.

At the border between stable nematic phase and unstable phases, flows induced by dipolar activity are just strong enough to trigger nematic deformations but not strong enough to nucleate topological defects and instead we observe the formation of \emph{extensile walls}. These extensile walls have also been observed previously~\cite{nejad2020memory} in wet active nematics in the presence of friction, and are a consequence of the competition between active forces to create defects and elastic forces to maintain the nematic phase. Going beyond the stability boundary, topological defects are nucleated in the system and we observe the emergence of several dynamic phases of defect organisation as a result of the competition between dipolar and quadrupolar active forces. 

To quantitatively distinguish between the stable phase (Fig~\ref{fig:lsa}, {\it yellow region}), where we either observe a nematic phase or extensile walls, and an unstable phase, where topological defects form and patterns emerge, we calculate the root-mean-square velocity averaged over both space and time after the statistical steady-state is reached. As evident from Fig.~\ref{rmsvel}a active flows decrease with increasing the quadrupolar force coefficient $\zeta_2$, which is consistent with the stabilising effect of the quadrupolar term.

The development of active turbulence stems from the onset of instabilities in the director field. Parallel walls consisting of lines of bend deformation separate the nematic regions. The decay of these walls leads to the formation of topological defects~\cite{thampi2014instabilities}. In order to quantify the nematic order and distinguish the nematic phase, we measure the average magnitude of the nematic order (defined as $S_0=\langle Q_{xx}^2+Q_{yx}^2 \rangle$). This quantity measures the global nematic order and thus indicates the overall stability of the nematic phase. A decrease in the value of the nematic order indicates the formation and presence of more defects. The average order is shown in Fig.~\ref{rmsvel}b for different values of the tumbling parameter. These graphs show the same trend as the average mean root square velocity graphs, indicating that flows are mainly created in places with large drop in the magnitude of the nematic order. Both graphs show a transition from the disordered phase to ordered phase by increasing the quadrupole force. 

 \begin{figure*}[t]
 \centering
    \includegraphics[width=1\textwidth]{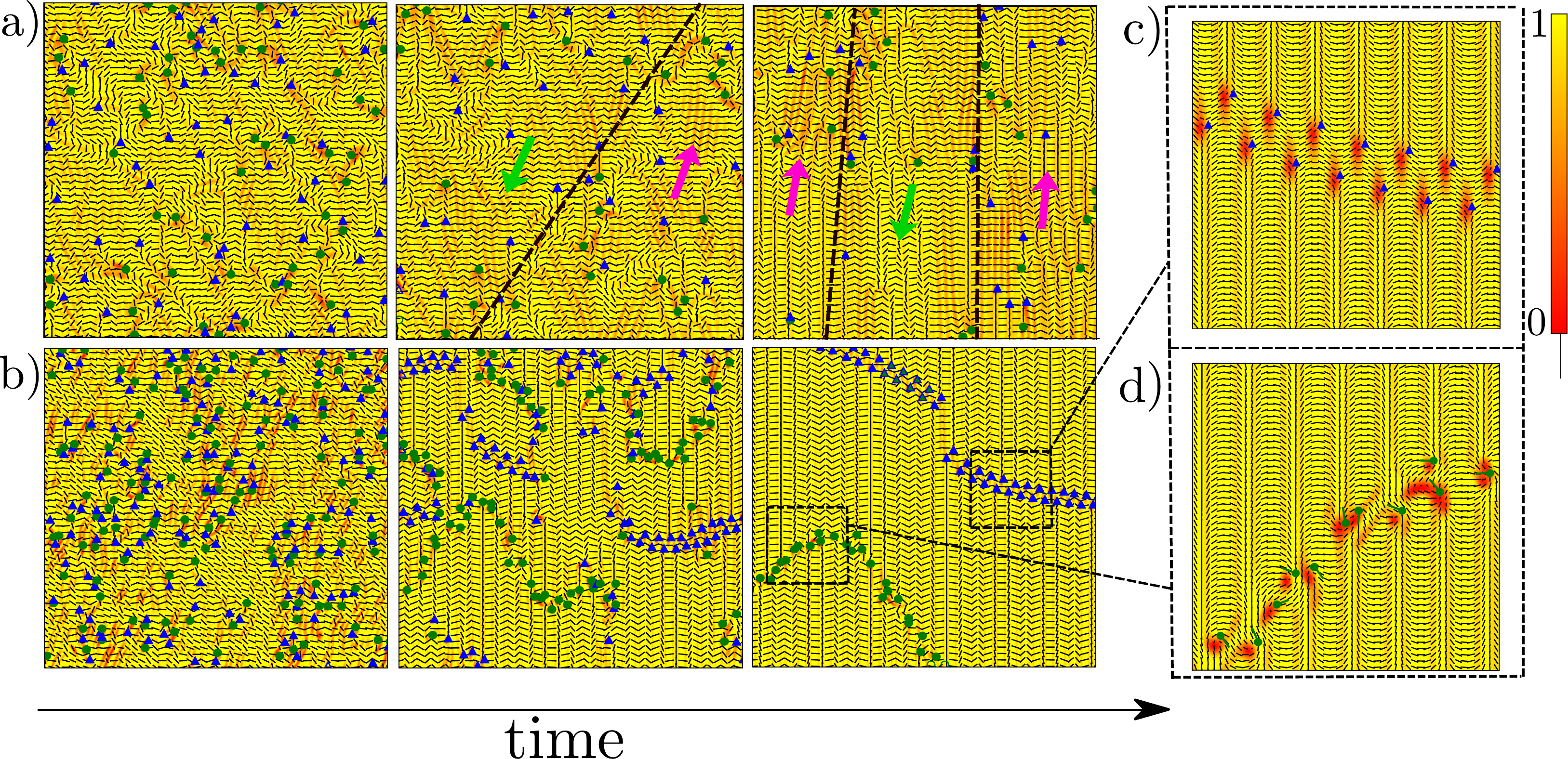}
    \caption{Time evolution of dynamics for (a) polar defect flock and (b) defect chains. In the polar defect flocking phase, the system forms motile defects that flock together in anti-parallel directions as indicated by the arrows. Defects in the chain phase self-organise to form (c) long chains of robust $-1/2$ defects as well as (d) chains of dynamic oscillating $+ 1/2$ defects as the system relaxes towards a steady state.}
\label{fig:snapshots}
 \end{figure*}

To further distinguish between the ordered and disordered phases of defect configurations we measure the average deformation, defined as $\langle\Delta n \rangle=\langle(\partial_i\partial_j Q_{ij})^2\rangle$, for different values of the tumbling parameter and as a function of activity ratio $\mathcal{Z}$. A large value of deformation follows a distortion in $\textbf{Q}$-tensor. Note that this quantity is different from the average order $S_0$ as it accounts for the distortions in the director and the deviation from the global nematic ordered phase. $\langle\Delta n \rangle$ is maximum in the presence of walls or in the active turbulence phase, as in these phases the global distortions in the director is maximum. The average distortion is presented in Fig.~\ref{deform} for varying values of the tumbling parameter. Note that ordered phases of defect organisation do not correspond to a peak in the distortion graph as in these phases the distortions in the director are local (in places of defects) and not global. 

Next, we describe the characteristics of defect self-organisation as this new non-linear regime is the focus of the current study.

\subsection{Dynamic phases}
We now turn to the unstable part of the phase diagram, where for a given value of the flow-aligning parameter, the instability driven by dipolar activity is strong enough to lead to deformation of the director field and nucleation of topological defects, while the presence of the quadrupolar activity acts to restructure the spatial defect organisation. In this regime, we found a variety of phases depending on the value of the tumbling parameter:
 
{\bf Defects flocking:} In the unstable phase, but close to the boundary of the stable phase, we observe a phase in which different groups of self-propelled $+1/2$ defects move together forming flocks of motile topological defects (Fig.~\ref{fig:lsa} and Supplementary Movie \hyperlink{https://www.dropbox.com/s/y5mto9519xvf9wo/1_pflock.mp4?dl=0}{$1$} in \cite{supplementary_movies}). Fig.~\ref{fig:angdist}a shows the angular distribution of $+1/2$ defects within this phase. Defects are identified using the diffusive charge density \cite{blow2014biphasic}
\begin{equation}
    s = \left(\frac{1}{2 \pi} \right) \times \left(\left[\frac{\partial Q_{xx}}{\partial x}\frac{\partial Q_{xy}}{\partial y} \right]
    - \left[\frac{\partial Q_{xx}}{\partial y}\frac{\partial Q_{xy}}{\partial x} \right] \right),\nonumber
\end{equation}
which takes values of $\pm 1/2$ at the defect cores and is zero otherwise. The angle $\theta$ is calculated using $\nabla \cdot \mathbf{Q} = (\cos \theta, \sin \theta)$, where $\nabla \cdot \mathbf{Q}$ gives the direction of the defects self-propulsion~\cite{pearce2021orientational} that for extensile systems ($\zeta_1 > 0$) is from tail to head of the $+1/2$ defect. Interestingly, in this phase, the defects in each flock move in the same direction and show polar order, but different defect flocks can migrate in anti-parallel directions within the system as seen from the snapshots in Fig.~\ref{fig:snapshots}a. 
To show this behavior more clearly, we calculate both polar and nematic order parameter for the alignment of the $+1/2$ topological defects. The polar ordering is defined by 
\begin{equation}
    P_{\text{+1/2}} = \sqrt{P_x^2+P_y^2},
\end{equation}
where 
\begin{equation}
    [P_x,P_y] = \left[\sum _i^N \frac{m_{xi}}{N} , \sum_i^N \frac{m_{yi}}{N} \right],
\end{equation}
and $m$ is defined by the orientational angle $\theta_i$ of each $+1/2$ defect as $m_{xi} = \cos\theta_i$ and $m_{yi} = \sin\theta_i$. Similarly, the nematic order of the defect alignment is characterised by the higher order multipole of the alignment angle \cite{DeGennes} 
\begin{equation}
    S_{\text{+1/2}} = \frac{1}{2} \langle 3   \cos^2 \theta -1 \rangle.
\end{equation}

For a perfect polar alignment of the motile defects $P_{\text{+1/2}} = 1.0$, while for the perfect nematic alignment $S_{\text{+1/2}} = 1.0$. For the angular distribution represented in Fig.~\ref{fig:angdist}a that corresponds to the defects flocking state, we find $P_{\text{+1/2}}=0.24$ and $S_{\text{+1/2}}=0.8$, indicating that the orientational organisation of the $+1/2$ defects in the defect flocks state shows a dominant nematic alignment.\\ 

{\bf Polar defect ordering:} Close to the defect flocking state, reducing the flow-aligning parameter together with lower quadrupolar activity leads to the global polar ordering of the motile $+1/2$ defects (Fig.~\ref{fig:lsa} and Supplementary Movie \hyperlink{https://www.dropbox.com/s/pn1w744ct4oqjp5/2_polar.mp4?dl=0}{$2$} in \cite{supplementary_movies}). This is reminiscent of the polar defect ordering predicted by the hydrodynamic theory of active defects in the over-damped limit and is due to the active aligning torque on defects in a charge neutral system~\cite{shankar2019hydrodynamics}. The polar ordering state is best represented in the angular distribution plot showing the sharp peak in the orientational alignment of the defects (Fig.~\ref{fig:angdist}b) and corresponds to the quantitative value of the global polar order $P_{\text{+1/2}}=0.5$.\\

{\bf Defect asters and active turbulence:} At the zero value of the flow-aligning parameter, where the director only responds to the rotational part of the flow gradient, and for negligible values of the quadrupolar activity coefficient, we find a different state characterised by aster-like organisation of groups of $+1/2$ defects (Fig.~\ref{fig:lsa} and Supplementary Movie \hyperlink{https://www.dropbox.com/s/99tdk44npgv0az9/3_aster.mp4?dl=0}{$3$} in \cite{supplementary_movies}). Increasing the strength of the quadrupolar activity towards negative values results in the breakup of the aster-like structures and the establishment of the active turbulence characterised by the chaotic motion of the topological defects. Within both phases the orientational organisation of the defects is isotropic (Fig.~\ref{fig:angdist}c,d and Supplementary Movie \hyperlink{https://www.dropbox.com/s/77t2ewndpvobl3u/4_turbulence.mp4?dl=0}{$4$} in \cite{supplementary_movies}).\\
    \begin{figure}[!ht]
\centering
    \includegraphics[width=\linewidth]{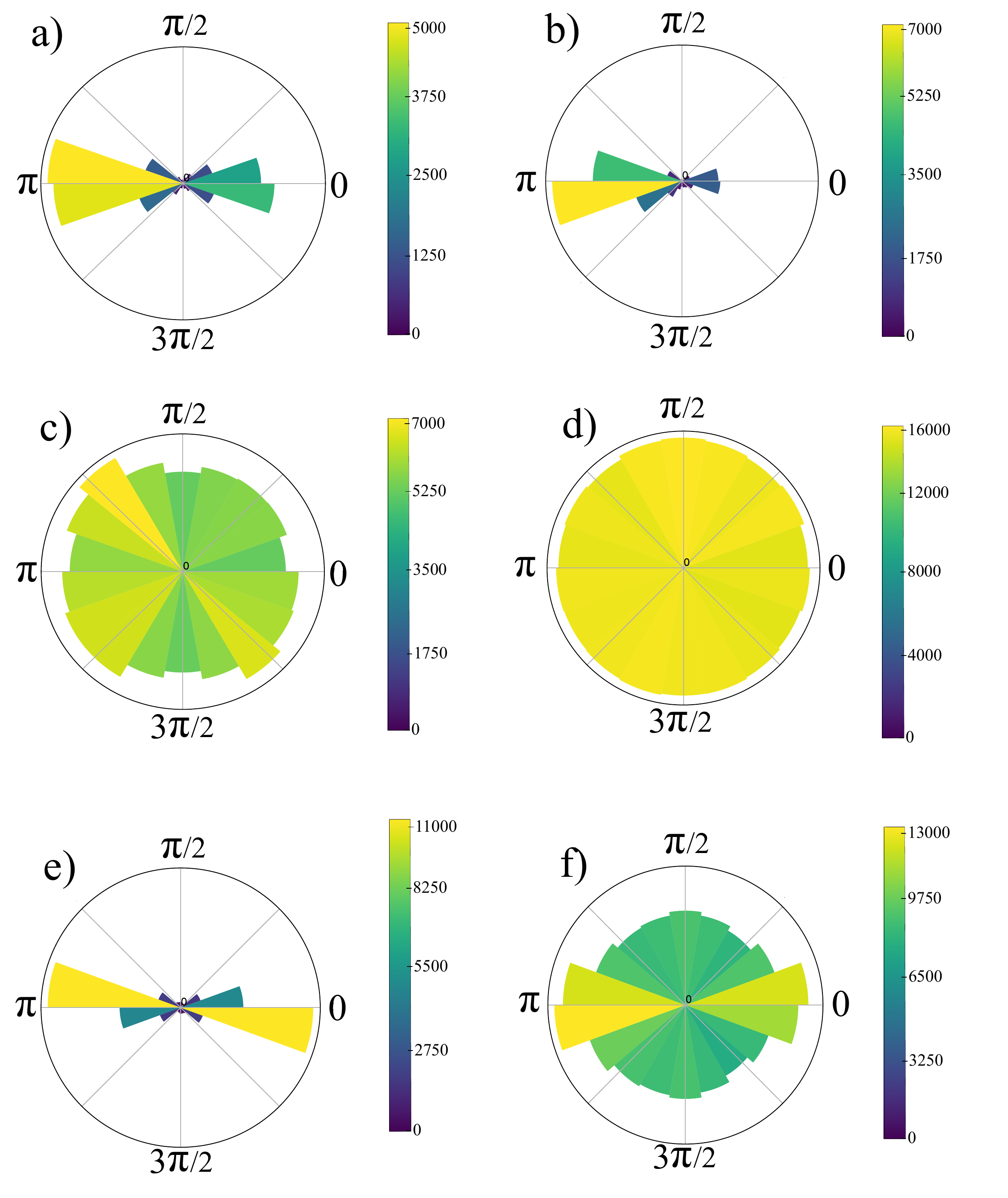}
\caption{Angular distribution of $+1/2$ defects for different phases: a) Defects flocking, b) Polar defect order, c) Active turbulence d) Defect asters, e) Defect chain and f) Polar defect flock coexisting with the defect chain. The different phases are distinguishable both by the direction of the $+1/2$ defects as well as the number of defects that form in a given phase. The colourbar corresponds to the number of $+1/2$ defects orientated at angle $\theta$
}
    \label{fig:angdist}
\end{figure}

{\bf Defect chains:} Interestingly, for  negative moderate quadrupolar activities, we found a phase in which $\pm 1/2$ defects self-organise into large chain-like structures. The chains of $+1/2$ and $-1/2$ defects show different characteristics: while $-1/2$ defect chains are stable and do not restructure after they are formed, the $+1/2$ defect chains are dynamic and within the chains pairs of $\pm 1/2$ defects nucleate, annihilate, move, and oscillate around each other (see Fig.~\ref{fig:snapshots}b and Supplementary Movie \hyperlink{https://www.dropbox.com/s/8i6qsrugk5pe280/5_chain.mp4?dl=0}{$5$} in \cite{supplementary_movies}) in a fashion resembling local defect dancing that has been reported for confined active nematics~\cite{doostmohammadi2019ceilidh}. The large-scale defect chains could be reminiscent of the propagating soliton-like clusters that have been recently discovered for confined active nematics with chiral anchoring~\cite{li2021soliton}, though here they are emergent properties of the system, where neither confinement nor chiral anchoring are present. In this vein, the impact of the friction on hydrodynamic screening and quadrupolar activity on breaking the angular momentum conservation could resemble the confinements and chiral anchoring effects, respectively, that have been shown to induce soliton-like defect clusters.

\begin{figure}[h]
 \centering
    \includegraphics[width=0.5\textwidth]{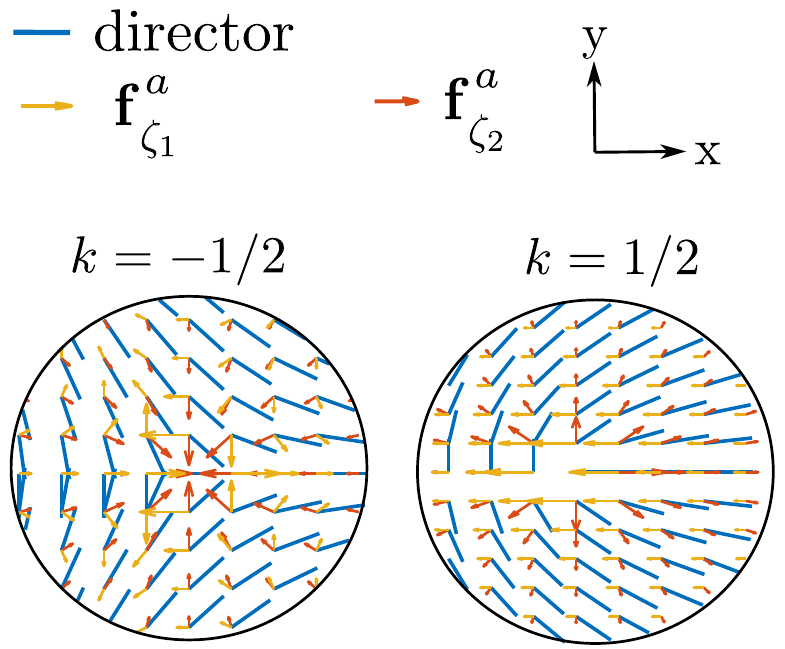}
    \caption{Force distribution around $\pm 1/2$ topological defects due to the dipolar and quadrupolar active forces, showing distinct diverging and converging force patterns due to the quadrupolar force around $+1/2$ and $-1/2$ defects, respectively. Here, blue solid lines indicate the director field of the defects, while yellow and red arrows illustrate force induced by dipolar and quadrupolar activities, respectively. Here, we have used $\mathcal{Z}=-0.75$.}
\label{fig:deflow}
 \end{figure}

In order to better understand the mechanism of chain formation we analytically calculate the force induced by dipolar and quadrupolar activities around isolated $\pm 1/2$ topological defects. The $\textbf{Q}$ tensor around a topological defect with charge $k$ reads: 
\begin{equation}
    \textbf{Q} =\begin{pmatrix}
\cos(2 k \phi) & \sin(2 k \phi)\\
-\sin(2 k \phi) & \cos(2 k \phi)
\end{pmatrix},
\end{equation}
where the defect symmetry axis is along the $x$-axis (as represented in Fig.~\ref{fig:deflow}), and $\phi$ is the polar angle (measured counter-clockwise with respect to the $x$-axis). 
Using this definition for the $\textbf{Q}$ tensor the active force around a defect reads:

\begin{equation}
\textbf{f}^a= \frac{q k}{r}
    \begin{pmatrix}
\zeta_2 \cos{\phi}-\zeta_1 \cos{\phi(1-2k)} \\
 \zeta_2 \sin{\phi+\zeta_1 \sin{\phi(1-2k)} }
\end{pmatrix}.
\end{equation}

From the symmetry of the quadrupolar active force it is expected that such a force does not have any impact on the self-propulsion speed of the defects. Interestingly, however, the quadrupolar activity induces distinct diverging and converging forces around $+1/2$ and $-1/2$ topological defects, respectively (Fig.~\ref{fig:deflow}). The converging force around an isolated $-1/2$ defect can explain the attraction of $-1/2$ defect pairs and the formation of chains of $-1/2$ defects (Fig.~\ref{fig:snapshots}b,c). On the other hand, the $+1/2$ defects that are in small distance from the $-1/2$ chain annihilate, but the $+1/2$ defects far from the $-1/2$ chain can form a dynamic chain. The orientation of the defects in a $+1/2$ chain, agrees with the active torque between $+1/2$ defects introduced in Ref. \cite{shankar2018defect} and suggests that accumulation of $-1/2$ defects in chains allows $+1/2$ defects to experience an aligning active torque purely caused by the other $+1/2$ defects. 

Finally, deep in the unstable phase we found a phase in which the polar defect flock phase coexists with the defect chain phase. In both pure defect chain phase and the coexistence phase, characterisation of the angular defect orientation reveals a dominant nematic alignment of the $+1/2$ defects (Fig.~\ref{fig:angdist}e, f).

\section{Conclusion}\label{sec3}
We have numerically studied the dynamics of active nematic systems in contact with a substrate. In this setup, a new active force contribution, that is absent in a bulk momentum conserving system, plays a role in the dynamics of the system \cite{maitra2018nonequilibrium}. We confirmed that for one sign of this active force, the nematic state becomes stable in agreement with the analytical result of Ref. \cite{maitra2018nonequilibrium}. Remarkably, going beyond linear stability, we also showed that in the unstable regime where the new active force cannot recover the nematic phase, it introduces new patterns of topological defect organisation in the system. In particular, we found the formation of stable chains of $-1/2$ topological defects, while the corresponding $+1/2$ defects form elongated clusters and oscillate around a center in pairs. By varying the tumbling parameter, we additionally found the emergence of polar defect flocks, in which system forms different flocks of $+1/2$ defects. Defects in each flock represent polar order but different flocks move in opposite direction and show nematic order. Furthermore, states showing pure polar flocking of the $+1/2$ defects and aster-like dynamic defect configurations, in which $+1/2$ defects point radially towards a center, were uncovered for varying quadrupolar-to-dipolar active force ratios and flow tumbling parameters.

This work could supplement further investigations into taming of active turbulence. Moreover, it provides numerical context to both the work in \cite{maitra2018nonequilibrium} as well as possible experimental assay results in which the quadrupolar force plays a role in the active stabilization or destabilization of the system. The exotic defect organisation phases revealed by the current numerical study can further trigger new experimental exploration of the topological defect structures in strongly confined active nematic systems. 

\section*{Acknowledgements}
M. R. N. acknowledges the support of the Clarendon Fund Scholarship. A. D. acknowledges support from the Novo Nordisk Foundation (grant No. NNF18SA0035142), Villum Fonden (Grant no. 29476), and funding from the European Union’s Horizon 2020 research and innovation program under the Marie Sklodowska-Curie grant agreement No. 847523 (INTERACTIONS).
\bibliographystyle{apsrev4-1}
\bibliography{ref.bib}
\end{document}